\documentclass{article}

\usepackage{graphicx}

\begin{document}

\title{Transmission Properties of Branched Atomic Wires}

\date{September 25, 2015}

\author{Kenneth W. Sulston\thanks{corresponding author} \thanks{Department of Mathematics and Statistics, University of Prince Edward Island, Charlottetown, PE, C1A 4P3, Canada}\and Sydney G. Davison\thanks{Department of Applied Mathematics, University of Waterloo, Waterloo, ON, N2L 3G1, Canada} \thanks{Department of Physics and the Guelph-Waterloo Physics Institute, University of Waterloo Campus, Waterloo, ON, N2L 3G1, Canada}}

\maketitle

\section{Abstract}

The renormalization-decimation method is used to study the transmittivity of atomic wires, with one
or more side branches attached at multiple sites. The rescaling process reduces all the branches,
attached at an atomic site, to an equivalent impurity, from which the transmission probability can be calculated
using the Lippmann-Schwinger equation. Numerical results show that the subsequent $T(E)$ curves, where
particular attention is paid to the numbers and locations of resonances and anti-resonances, are highly sensitive
to the values of each system's key parameters. These findings provide insight into the design of
wires with specific desired properties.

\section{Introduction}

Modern fabrication techniques allow electronic devices to be designed and constructed at the molecular
level. Consequently, it is instructive to investigate possible structures theoretically, with a view to their
varying physical properties, as a guide to designing devices with the desired characteristics. Of interest here
are the electronic transmission aspects of the atomic wire, and how they can be modified by the
attachment of one or more atoms (or groups of atoms) at various points on the wire. 

An early study, along these lines, was that of Guinea and Verg\'{e}s \cite{ref1}, who investigated the 
electronic properties of a chain with linear side branches and loops. The work concentrated on the local
densities of states, but also considered the implications for transmission. A key finding was that the
irregularities in geometric structure lead to the blocking of transmission, at certain energies. A later
treatment, focusing on transmisison through a chain, was performed by Singha Deo and Basu \cite{ref2}.They showed
that an interplay between substitutional impurities and topological defects (such as a side branch)  can make the
band asymmetric, with resulting effects on the transmission. The effect can arise with as little as a single
impurity present in an otherwise periodic chain. More recently, Kalyanaraman and Evans \cite{ref3}
also considered electron transfer along an atomic wire with a side branch, and with particular attention paid
to dendritic structures. The main result was that such structures can exhibit larger conductance than linear
chains, with the position and topology of the side branches being crucial in determining the transport properties.
Farchioni {\em et al} \cite{ref4} utilised the renormalization method (see, e.g., \cite{ref5}) to study
electron transmission through a ladder (two coupled chains) with a side-attached impurity, observing that
the effects are quite different than for a substitutional impurity.

The present authors \cite{ref6} used a tensorial Green-function method to study overlap effects on 
electron transmission through the simplest type of branched
atomic wire, namely, a T-junction consisting of a chain of atoms with just a single additional atom attached
to its side. It was observed that overlap can either enhance or suppress transmission, and does so in a
different way than does the presence of an impurity or a branch in the chain.

The present paper extends the previous work \cite{ref6} to consider more elaborate attachments, specifically,
longer branches with sub-branches and multiple branches at a single site, and branches at two sites along
the atomic wire. The method used is the tensorial Green-function technique of earlier work \cite{ref6, ref7}, which
allows for the inclusion of overlap both in the atomic wire and any attached structures. The renormalization
method \cite{ref5} provides an efficient way to treat a complicated attachment, by rescaling it to an 
energy-dependent substitutional impurity at the chain site to which the structure is attached. As we are
considering such attachments at either one or two sites, such systems can thus be reduced to atomic
wires with one or two impurities. As previously mentioned, the former has been considered by us
\cite{ref7}, while the latter has only been treated in the zero-overlap case \cite{ref8}. Thus, the methodology,
discussed in the next sections, firstly extends the theory of electron transmission through double-impurity wires,
so as to incorporate overlap, which is similar to the single-impurity theory, then, secondly, uses the renormalization
technique to reduce complicated branched structures to single atomic wires containing one or two impurities.

\section{Transmission through Double-impurity Wires, including Overlap}

\begin{figure}[h]
\includegraphics[width=12cm]{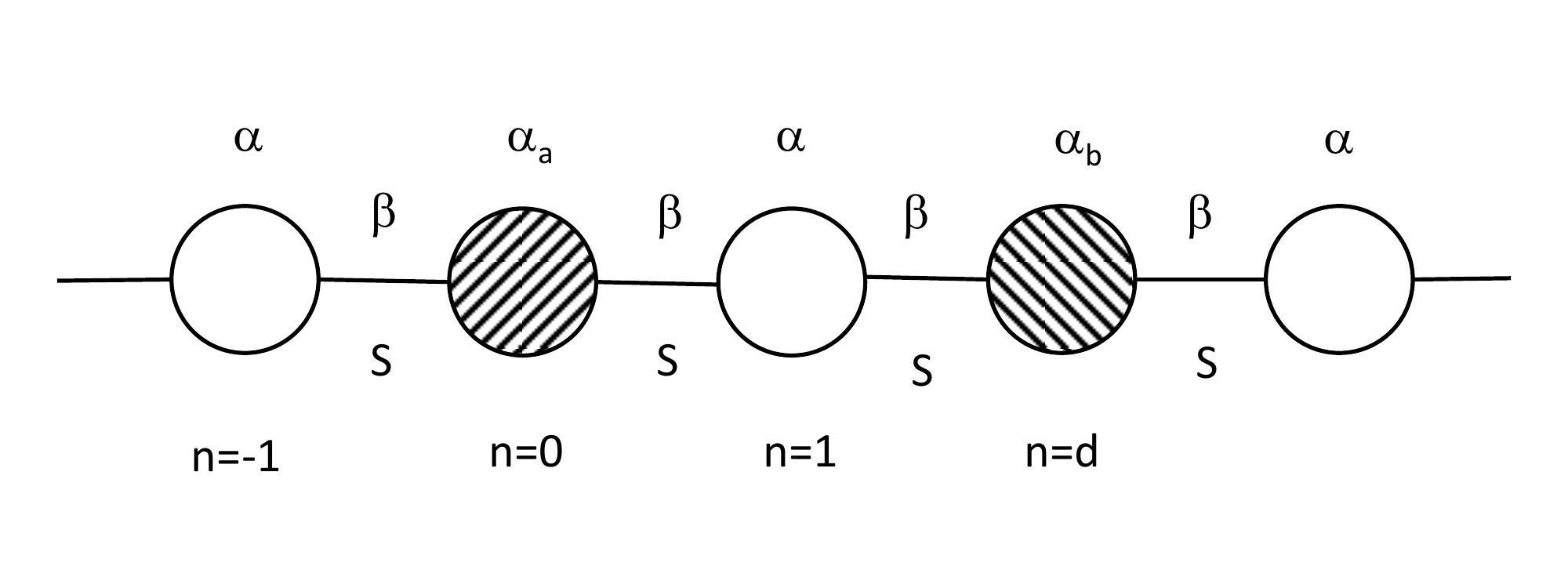}
\caption{Atomic wire with 2 impurities, at separation $d$.}
\label{fig1}
\end{figure}

In this section, we look briefly at the theory of electron transmission through an atomic wire,
with two impurities at a separation $d$, as shown in Figure \ref{fig1}. 
The wire is modelled
by a linear chain of atoms, with $\alpha$ being the atomic-site energy, $\beta$ the bond energy,
and $S$ the nearest-neighbour overlap. The impurities, at the sites $n=0$ and $n=d$, have modified
site energies $\alpha_a$ and $\alpha_b$, respectively, with the obvious feature that the system
reduces to a one-impurity chain by setting $\alpha_b = \alpha$. Transmission through such a
wire for the case of no overlap ($S=0$) has been treated by Mi\v{s}kovi\'{c} {\em et al} \cite{ref8},
from which the extension to the case $S \ne 0$ is straightforward, using the method of reference \cite{ref7}.
To this end, with some changes in notation, the Mi\v{s}kovi\'{c}  equation (21) for the
transmission probability becomes
\begin{equation}
T(E) = | \tau |^2 = |(1-i c_a)(1-i c_b)+c_a c_b t^{2d}|^{-2} ,
\label{eq1}
\end{equation}
where
\begin{equation}
c_k = s (\alpha_k - \alpha )(1-\xi^2)^{-1/2}  / (4 \beta Y) ~,~ k = a ~{\rm or}~ b,
\label{eq2}
\end{equation}
and
\begin{equation}
t =  \left\{ \begin{array}{ll}
  \xi+s(\xi^2-1)^{1/2} , &  |\xi| > 1 , \\
  \xi+is(1-\xi^2)^{1/2} , & |\xi| < 1 ,
    \\
    \end{array}
  \right.  
\label{eq3}
\end{equation}
with
\begin{equation}
s = \pm 1 ~{\rm so~ that}~ |t| < 1.
\label{eq4}
\end{equation}
Here,
\begin{equation}
\xi = X/2Y = (E- \alpha) / 2 (\beta - ES) , 
\label{eq5}
\end{equation}
where
\begin{equation}
X = (E - \alpha)/ 2 \beta ,
\label{eq6}
\end{equation}
and
\begin{equation}
Y = (1 - \alpha S/ \beta - 2SX)/2 .
\label{eq7}
\end{equation}
The fundamental mathematical change in including overlap is that the bond energy $\beta$
is replaced by the energy-dependent quantity $\beta - E S$. Conversely, if overlap is removed
by setting $S=0$, then the above results reduce to those of \cite{ref7}.

\section{Renormalization Technique}

\begin{figure}[htpb]
\includegraphics[width=12cm]{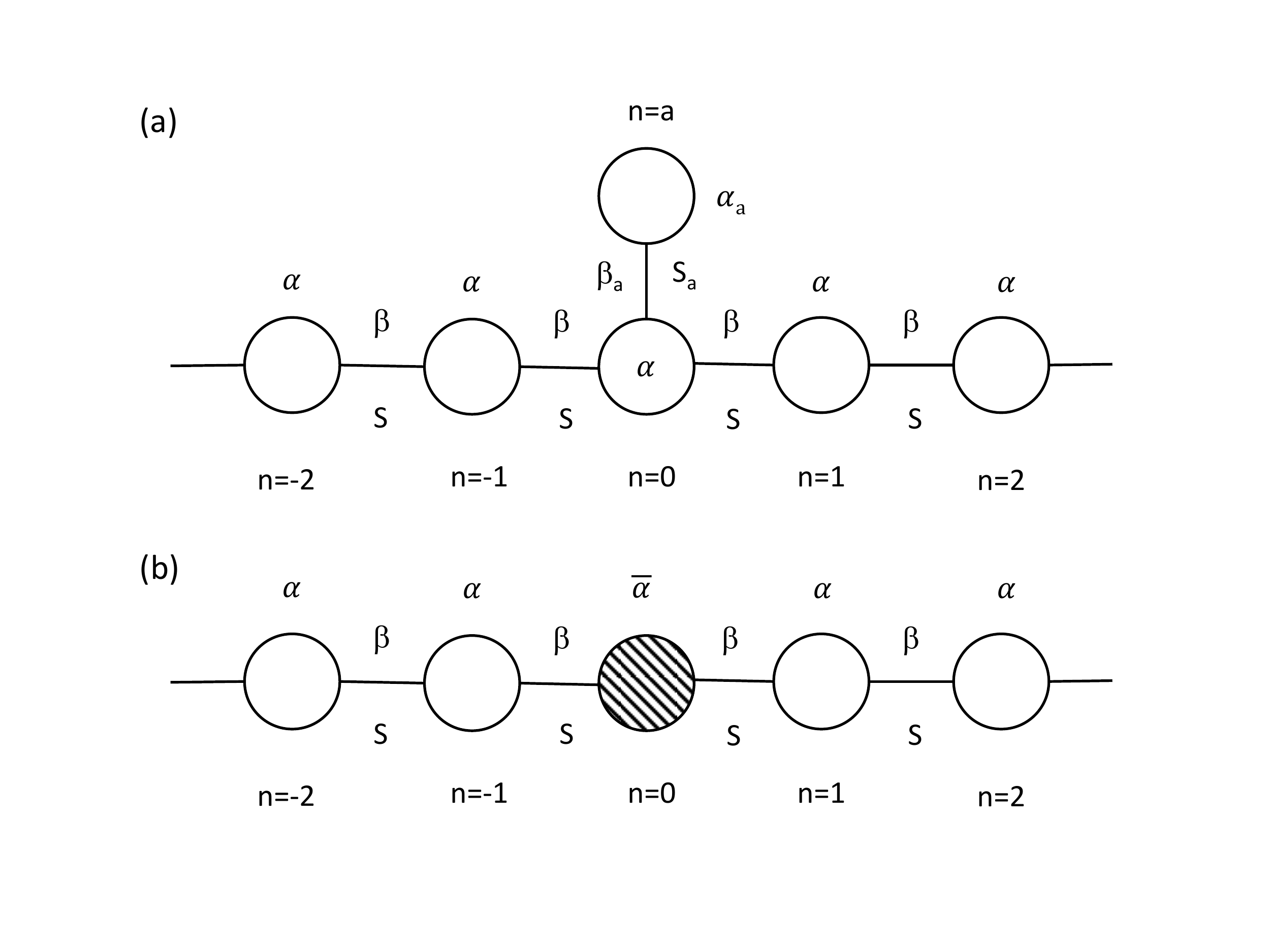}
\caption{(a) Chain with one side-atom attached. (b) Rescaled chain with side-atom deleted.}
\label{fig2}
\end{figure}

The renormalization technique \cite{ref5} provides an efficient avenue to investigate
complicated structures, by rescaling them into simpler ones, with one or more
parameters becoming energy-dependent.
Although the method is most commonly used without inclusion of overlap, it is
straightforward to do so. We here illustrate the method by looking at its
application to the simplest system of interest in this paper, namely a linear
chain with a single extra atom attached at the side of site 0 (see Figure \ref{fig2}(a)). 
The pure chain is as described in the last section, while the side-atom has a site energy
$\alpha_a$, bond energy $\beta_a$ and overlap $S_a$. Within the tight-binding
approximation, the discretized Schr\"{o}dinger equation takes the well-known form \cite{ref7}
of a set of second-order difference equations, which for the system under consideration is:
\begin{equation}
(E - \alpha ) c_{n} = (\beta - ES) (c_{n-1} + c_{n+1}) ~,~ n \ne 0,a ,
\label{eq8}
\end{equation}
\begin{equation}
(E - \alpha ) c_{0} = (\beta - ES) (c_{-1} + c_{1}) +  (\beta_a - ES_a) c_{a},
\label{eq9}
\end{equation}
\begin{equation}
(E - \alpha_a ) c_{a} = (\beta_a - ES_a) c_{0}.
\label{eq10}
\end{equation}
Renormalization proceeds by eliminating $c_a$ from the set. Equation (\ref{eq10})
can be solved as
\begin{equation}
c_{a} = {(\beta_a - ES_a) \over (E - \alpha_a )}  c_{0},
\label{eq11}
\end{equation}
which, when substituted into (\ref{eq9}), produces
\begin{equation}
(E - \bar{\alpha} ) c_{0} = (\beta - ES) (c_{-1} + c_{1}) ,
\label{eq12}
\end{equation}
where
\begin{equation}
\bar{\alpha} = \alpha + {(\beta_a - ES_a)^2 \over (E - \alpha_a )}  
\label{eq13}
\end{equation}
is the renormalized site energy. Thus, equation (\ref{eq10}) has been
eliminated, leaving the system of equations as (\ref{eq8}) along with
(\ref{eq12}), and represented schematically by Figure \ref{fig2}(b).
The procedure can be extended to more complicated systems, as will
be seen in the next section, by eliminating sites, one-by-one, starting at
the end of a branch, which corresponds mathematically to recursive
application of (\ref{eq13}) to produce the renormalized site energy at
the ``root'' site, which appears as an impurity in the chain.

\section{Application to Branched Atomic Wires}

\subsection{Chain with Branch and Sub-branch}

\begin{figure}[htpb]
\includegraphics[width=12cm]{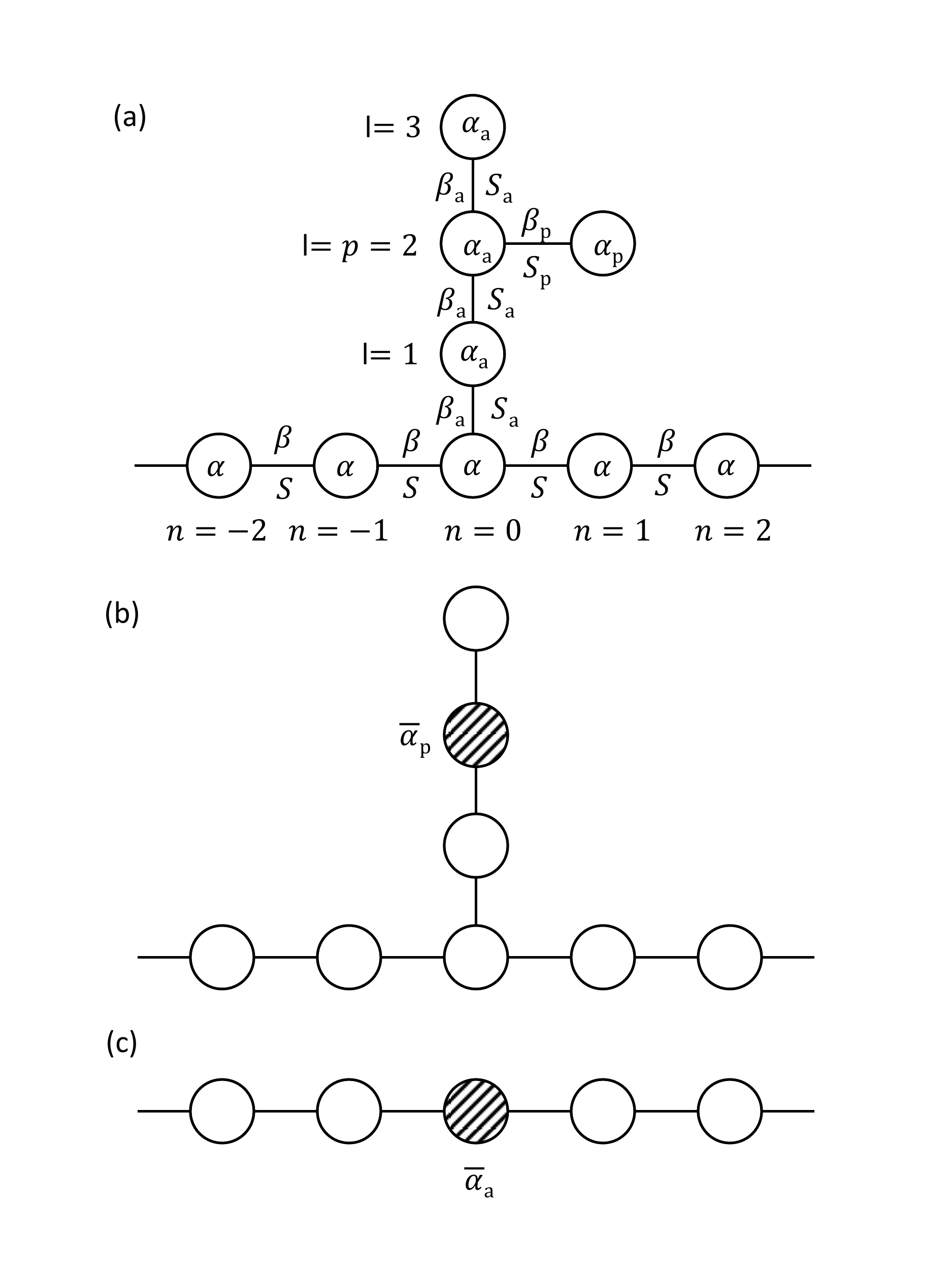}
\caption{(a) Chain with one branch plus side-atom. (b) Result of first renormalization. (c) Final result of
renormalization process.}
\label{fig3}
\end{figure}

We now proceed to investigate electron transmission through certain branched atomic-wire systems,
by using the renormalization method to reduce such systems to linear chains with one or two impurities,
whose transmission is then given by (\ref{eq1}). The first system under study is an atomic wire with
a single side branch, which itself has a sub-branch (see Figure \ref{fig3}(a)). 
The branch is of
length $l_a$, and is attached to the main wire at site $n=0$. Each atom in the side branch has site
energy $\alpha_a$, bond energy $\beta_a$ and overlap $S_a$. The sub-branch is attached to the
primary branch at site $l=p$, and, for simplicity, the sub-branch is taken to consist of just a
single atom, although the generalization to a longer sub-branch is straightforward. The atom comprising
the sub-branch is taken to have site energy $\alpha_p$, bond energy $\beta_p$ and overlap $S_p$.
The first step in the renormalization process is to delete the sub-branch atom, resulting in a rescaled
atom at site $l=p$  (Figure \ref{fig3}(b)), whose site energy, using (\ref{eq13}), is
\begin{equation}
\bar{\alpha}_p = \alpha_a + {(\beta_p - ES_p)^2 \over (E - \alpha_p )}   .
\label{eq14}
\end{equation}
The second step is to decimate the branch, atom by atom, starting at the free end ($l=l_a$),
through repeated application of (\ref{eq13}), resulting in only a chain with a rescaled atom at site
$n=0$ (Figure \ref{fig3}(c)), whose site energy can be written in continued-fraction form as
\begin{equation}
\bar{\alpha}_a = \alpha + {(\beta_a - ES_a)^2 \over E - \alpha_a - }  ~ {(\beta_a - ES_a)^2 \over E - \alpha_a - } \cdots
 {(\beta_a - ES_a)^2 \over E - \bar{\alpha}_p - }  \cdots {(\beta_a - ES_a)^2 \over E - \alpha_a } ,
\label{eq15}
\end{equation}
where the term $\bar{\alpha}_p$ occurs at position $p$ (counting from the left end of the continued
fraction). Alternatively, (\ref{eq15}) can be expressed more compactly using Gauss' notation as
\begin{equation}
\bar{\alpha}_a = \alpha -{\rm K}_{l=1}^{l=l_a} \Big[ {-(\beta_a - ES_a)^2 \over E - \alpha_l } \Big]   ,
\label{eq15a}
\end{equation}
where
\begin{equation}
\alpha_l =  \left\{ \begin{array}{ll}
  \alpha_a , &  l \ne p , \\
  \bar{\alpha}_p , & l=p .
    \\
    \end{array}
  \right.  
\label{eq15b}
\end{equation}
Hence, the transmission can be calculated as that through a 1-impurity wire, by using equations
(\ref{eq1})-(\ref{eq7}), with $\alpha_a \rightarrow \bar{\alpha}_a$ and  $\alpha_b \rightarrow \alpha$.

\subsection{Chain with Multiple Branches at One Site}

\begin{figure}[htpb]
\includegraphics[width=12cm]{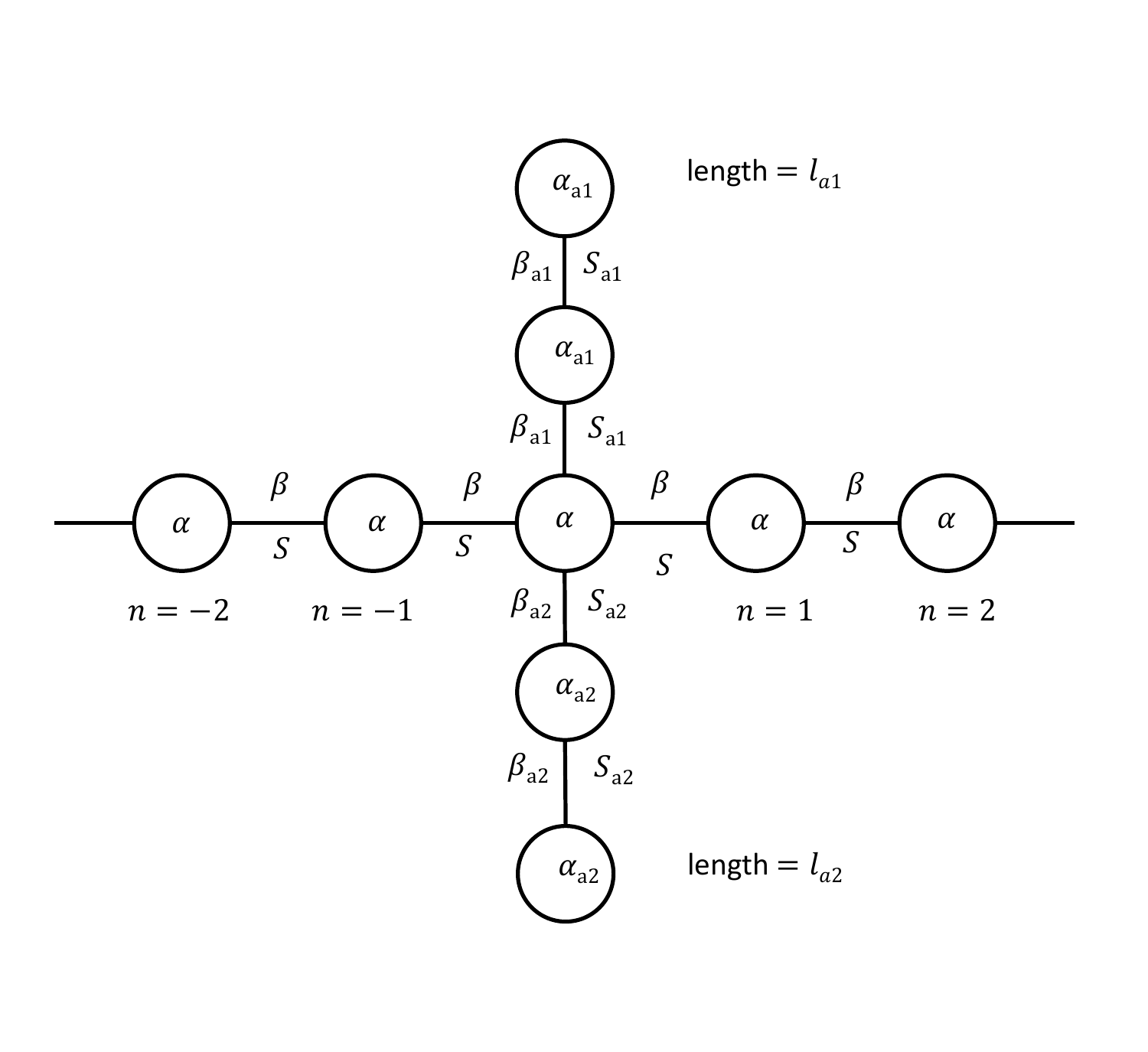}
\caption{Chain with side branches at one site.}
\label{fig4}
\end{figure}

Next, we turn to the system of an atomic wire with $M$ branches attached at the same site ($n=0$),
as shown in Figure \ref{fig4}.
On branch $m$ ($=1,...,M$), which is of length $l_m$,
the atoms have site energy $\alpha_{am}$, bond energy $\beta_{am}$ and overlap $S_{am}$.
Each branch can be decimated via the renormalization procedure, as was
done in the previous subsection, albeit without sub-branches attached. Consequently, each branch contributes
to $\bar{\alpha}_a$ a continued fraction of the form appearing in (\ref{eq15a}), resulting in the rescaled
site-energy at site $n=0$ being
\begin{equation}
\bar{\alpha}_a = \alpha - {\textstyle\sum_{m=1}^{m=M}}  \Big( {\rm K}_{l=1}^{l=l_m} \Big[ {- (\beta_{am} - ES_{am})^2 \over E - \alpha_{am}} \Big]      \   \Big)  .
\label{eq16}
\end{equation}
Thus, as in the last subsection, the transmission is through a rescaled 1-impurity wire, with
$\alpha_a \rightarrow \bar{\alpha}_a$ and  $\alpha_b \rightarrow \alpha$.

\subsection{Chain with Branches at Two Sites}

\begin{figure}[htpb]
\includegraphics[width=12cm]{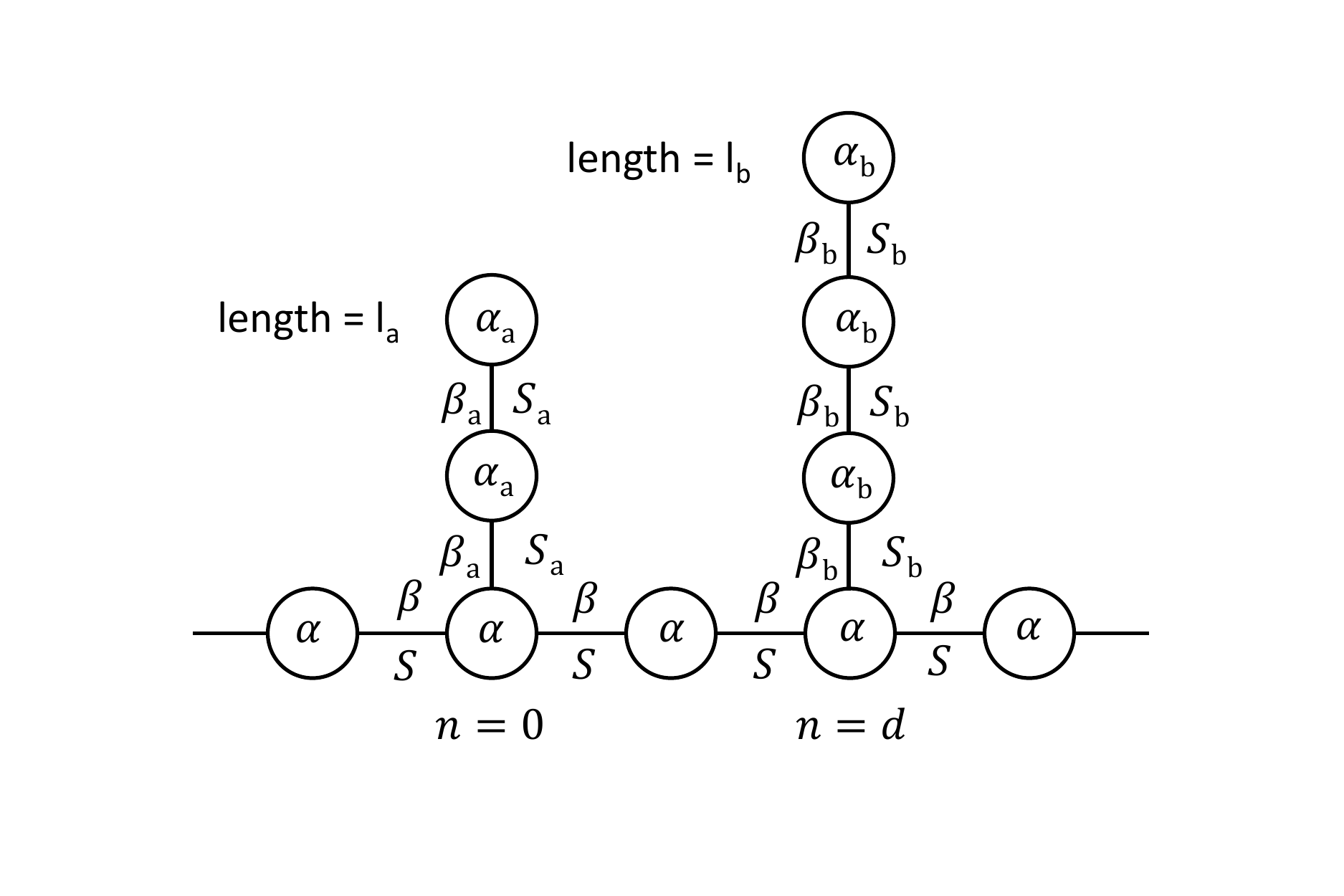}
\caption{Chain with side branches at two sites, at separation $d$.}
\label{fig5}
\end{figure}

The last system under investigation is an atomic wire, with a pair of branches attached at
sites $n=0$ and $n=d$ (see Figure \ref{fig5}). 
The branches are of lengths $l_a$ and $l_b$, 
respectively, and have associated site energies $\alpha_k$, bond energies $\beta_k$ and overlaps $S_k$,
with $k = a~{\rm or}~b$. Each branch can be decimated, atom by atom, as in the preceding subsections
by the repeated application of (\ref{eq13}). Consequently, sites 0 and $d$ of the chain become occupied
by rescaled atoms with site energies
\begin{equation}
\bar{\alpha}_k = \alpha -{\rm K}_{l=1}^{l=l_k} \Big[ {-(\beta_k - ES_k)^2 \over E - \alpha_k } \Big]   ,~k=a ~{\rm or}~b,
\label{eq17}
\end{equation}
respectively. Hence, the transmission $T(E)$ can be calculated, via (\ref{eq1}), as that through a 2-impurity wire, with
$\alpha_a \rightarrow \bar{\alpha}_a$ and  $\alpha_b \rightarrow \bar{\alpha}_b$.

\section{Results and Discussion}

We now proceed to examine the $T(E)$ curves for the three systems, for some
specific values of the various parameters. In order to provide some degree of 
uniformity, we take as ``standard'' values, all site energies $\alpha$'s to be 0,
bond energies $\beta$'s to be -0.5 and overlaps $S$'s to be 0.25, except as 
otherwise stated. 

\begin{figure}[h]
\includegraphics[width=12cm]{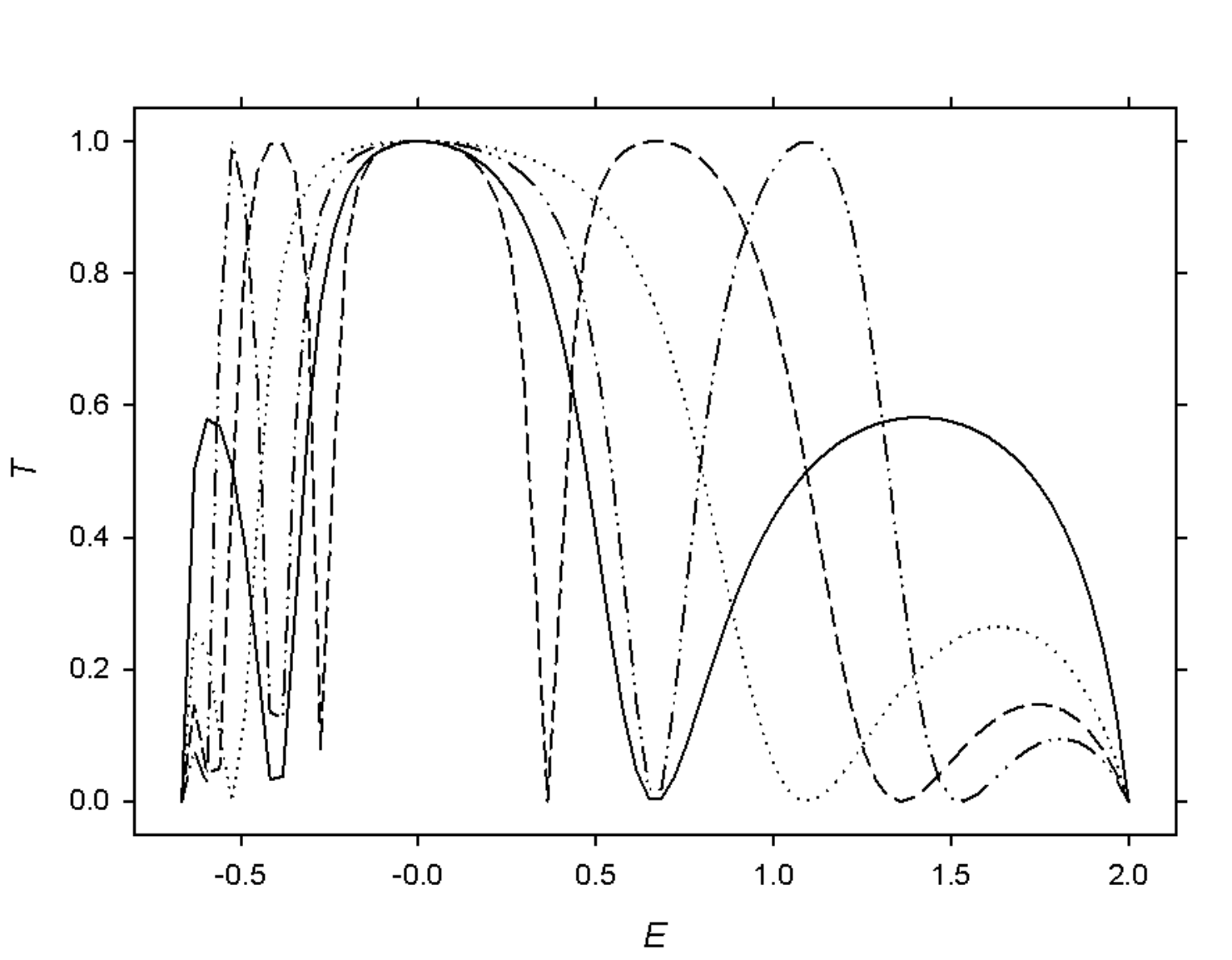}
\caption{$T(E)$ for chain with branch, plus side-atom attached at position $p=1$. The branch is of length $l_a=$
(a) 1 (solid curve), (b) 2 (dotted), (c) 3 (dashed), (d) 4 (dash-dotted).}
\label{fig6}
\end{figure}

Turning first to the system of a chain with one branch plus a side-atom, the
key parameters to investigate are the length of the branch $l_a$ and the postion $p$ of
the side-atom on that branch. With respect to the former, Figure \ref{fig6} shows
the $T(E)$ graphs for several branch lengths, with the side-atom set at position $p=1$,
i.e., attached to the branch-atom closest to the main chain. 
Starting with the case $l_a=1$
(Figure \ref{fig6}(a)), we see a single resonance ($T=1$) at $E=0$, flanked by a pair of
anti-resonances ($T=0$), which separate the central peak from two smaller ones nearer
the band edges. By comparison, $T(E)$ for a chain with a branch, but no side-atom, shows one resonance
and just a single anti-resonance (see Figure 2 of \cite{ref6}), while that for a bare chain, with no branch, shows
only the single resonance and no anti-resonances at all (see Figure 5 of \cite{ref7}). Moving to a
branch of length $l_a=2$ (Figure \ref{fig6}(b)), we see a very similar $T(E)$ curve, namely, a large
central peak with resonance at $E=0$, with 2 smaller peaks at the sides, albeit with the 2 anti-resonances
shifted somewhat outwards, towards the band edges. However, further lengthening of the branch to
$l_a=3$ (Figure \ref{fig6}(c)) produces a very different $T(E)$ curve, in which the central peak has split
into 3, each achieving a resonance, while the 2 outermost peaks are further diminished, and also narrowed,
by virtue of their bounding anti-resonances again moving further outwards. These features are repeated
in the next longer branch ($l_a=4$ in Figure \ref{fig6}(d)), but once more with all anti-resonances moving 
outwards. This pattern is repeated as the branch is further lengthened (graphs not shown), with pairs
of graphs for branch lengths $2m-1$ and $2m$ appearing very similar, but with an additional pair of anti-resonances
appearing, due to a split of the central peak, in going to lengths $2m+1$ and $2m+2$.

\begin{figure}[h]
\includegraphics[width=12cm]{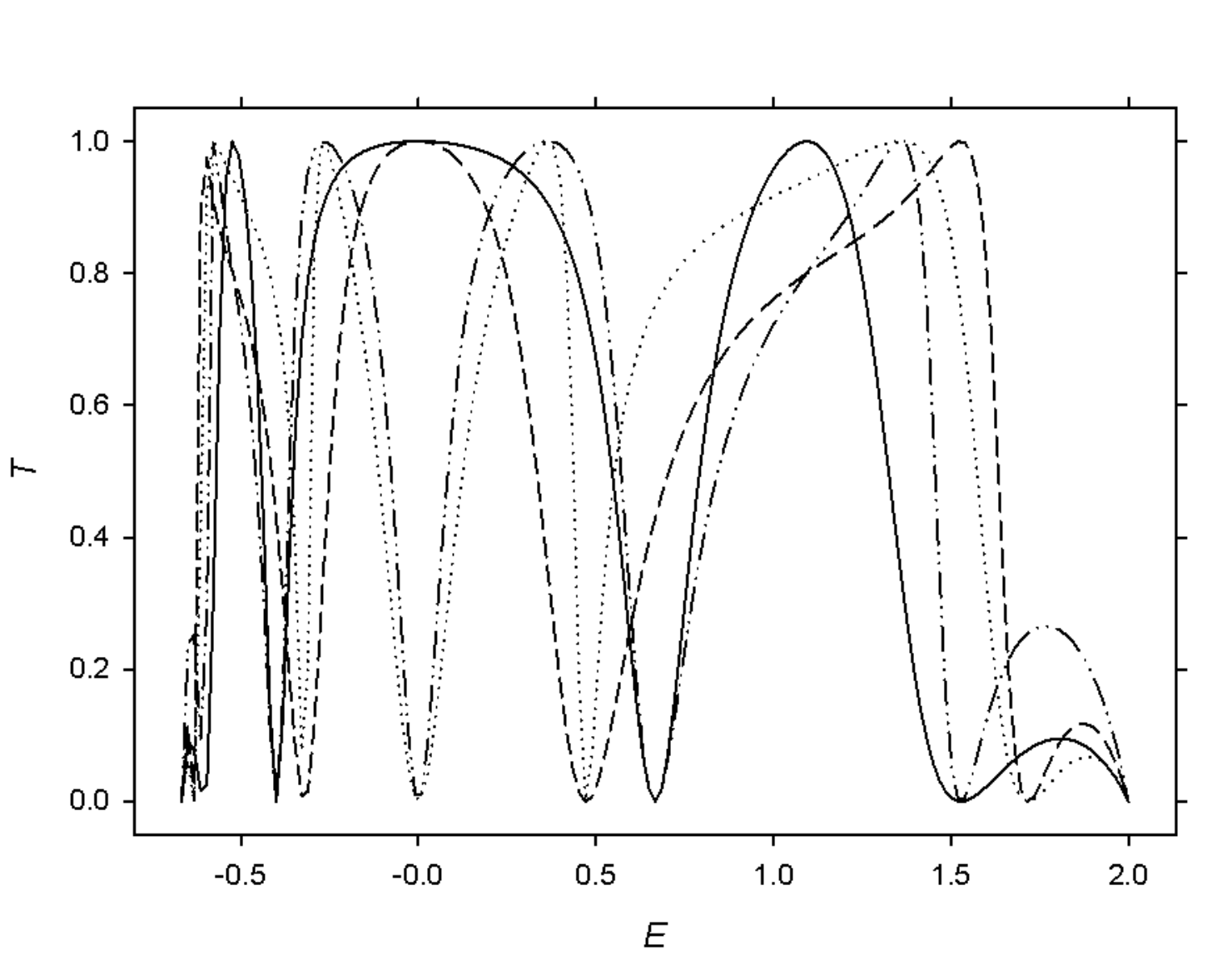}
\caption{$T(E)$ for chain with branch of length $l_a=4$, plus side-atom attached at position $p=$
(a) 1 (solid curve), (b) 2 (dotted), (c) 3 (dashed), (d) 4 (dash-dotted).}
\label{fig7}
\end{figure}

The dependence of the $T(E)$ curves, on the position $p$ of the side-atom on the
branch, is illustrated in Figure \ref{fig7}, for a branch of length $l_a=4$. 
The graph 
for $p=1$ (Figure \ref{fig7}(a)), it should be noted, is that appearing in Figure \ref{fig6}(d),
again showing the structure of 5 peaks, 3 of them resonances, divided by 4 anti-resonances.
When the side-atom is moved outwards along the branch to position $p=2$ (Figure \ref{fig7}(b)),
the central resonance, centered at $E=0$, is split into a pair of narrower resonances, accompanied
by some shifting of the anti-resonances, so as to broaden the other 2 resonances. As the side-atom
is moved again to $p=3$ (Figure \ref{fig7}(c)), however, these 2 new resonances recombine into a
single one at $E=0$, with the 2 outer resonances relatively unchanged. The graph now looks very
similar to that for $p=1$ in Figure \ref{fig7}(a), in terms of number of resonances and anti-resonances,
except that the central resonance peak has been narrowed, allowing broadening of the outer two. With
an increase to $p=4$  (Figure \ref{fig7}(d)), the central resonance agains splits into a pair of resonances.
This pattern repeats itself, on branches of arbitrary length, as $p$ increases, namely, the number of
resonances oscillates as the central resonance splits in two and then recombines.

A detailed mathematical analysis can reveal much about the features seen in Figures \ref{fig6}
and  \ref{fig7} and, in particular, about the dependance on the parameters of the number of
anti-resonances, which largely determines the overall  structure of the graphs. Referring to
equation (\ref{eq1}), it can be seen that in order for an anti-resonance to occur, i.e., $T=0$,
it is necessary that $c_l \rightarrow \pm \infty$. Looking at  (\ref{eq2}),  
$\alpha_b \rightarrow \alpha$ (as indicated in section 5.1)  giving $c_b=0$, thus leaving
only $c_a$ to consider. For $c_a \rightarrow \pm \infty$ in (\ref{eq2}) requires that 
$\bar{\alpha}_a - \alpha \rightarrow \pm \infty$, as it can be shown that $Y \ne 0$ and
$1 - \xi^2 =0$ only at the band edges (where T always equals 0, so these two energies are not
considered to be anti-resonances). Thus, anti-resonances correspond to singularities of 
$\bar{\alpha}_a - \alpha$, which can be located by analyzing  (\ref{eq15a}), with the 
result that the maximum, and ``usual'', number of anti-resonances equals $l_a+1$, namely,
 the total number of atomic sites in the entire side-structure (branch plus side-atom). However,
for some parameter values, symmetry considerations can cause a coalescence of 2 anti-resonances,
in a situation reminiscent of degeneracy.  Suppose $\alpha_a = \alpha_p$, as is the case here.
If, for example, $p=l_a -1$, then the side-structure has a symmetrical Y-shape, which results in the
coalescence of 2 anti-resonances, thus dropping the total number to just $l_a$. This feature does
not happen when $p=l_a$ or $l_a-2$. It depends upon the parity (even or odd) of $p$ compared
to that of $l_a$, and it turns out that the number of anti-resonances is $l_a+1$ when $p$ and $l_a$
have the same parity, but the number drops to $l_a$, when they have opposite parity.
It should be noted that, if $\alpha_a \ne \alpha_p$, then the symmetry consideration does not
hold, and the number of anti-resonances is $l_a +1$.
These considerations explain the basic structure of the two figures. Firstly, in Figure \ref{fig6},
increasing $l_a$, while holding $p$ constant, switches the parity back and forth, resulting in pairs
of graphs, with branch lengths $2m-1$ and $2m$, having the same number of anti-resonances.
Secondly, in Figure \ref{fig7}, increasing $p$, while holding $l_a$ constant, also flips the parity, 
causing the number of anti-resonances to oscillate up and down.

\begin{figure}[h]
\includegraphics[width=12cm]{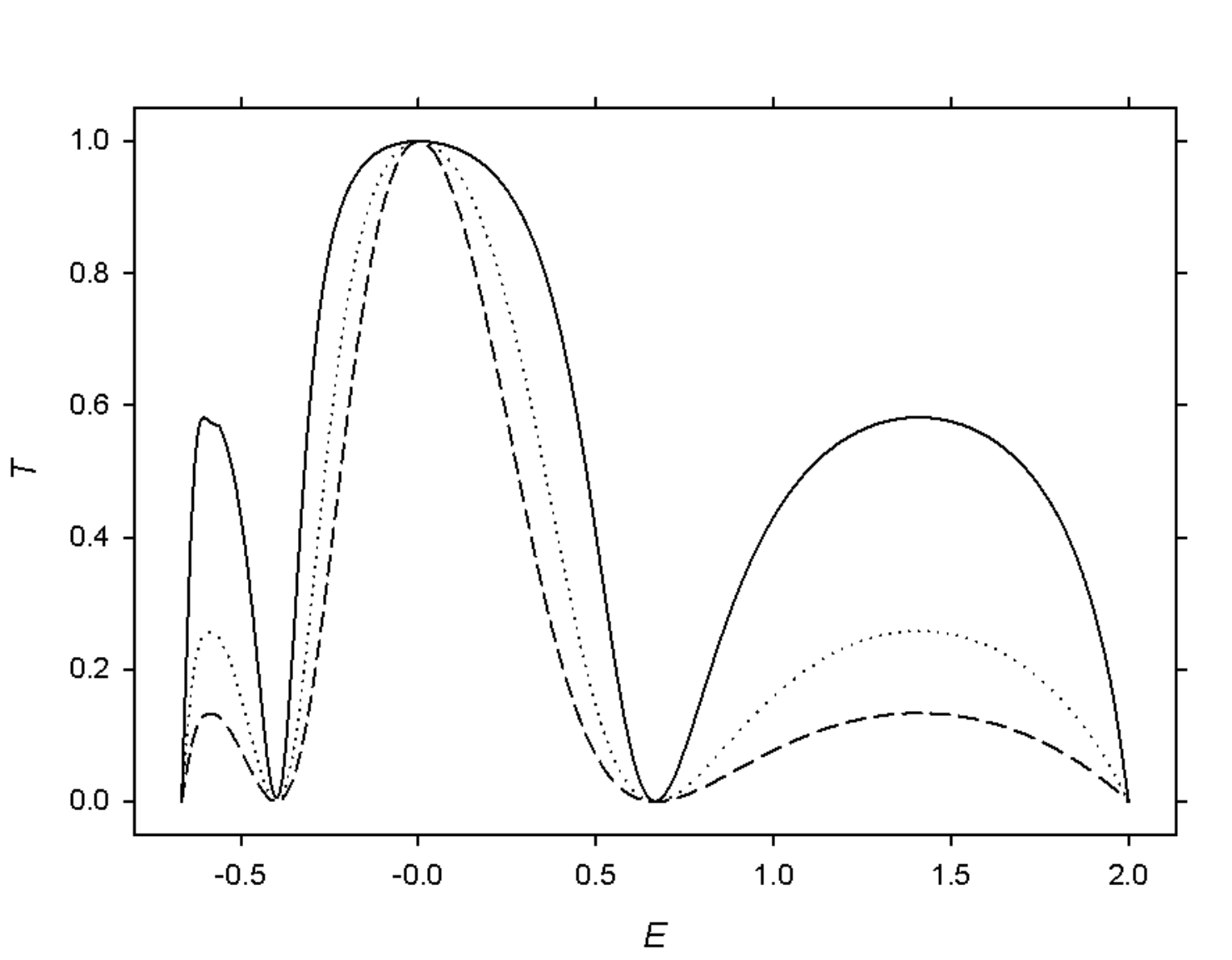}
\caption{$T(E)$ for chain with branches of length $l_m=2$, the number of branches being $M=$
(a) 1 (solid curve), (b) 2 (dotted), (c) 3 (dashed).}
\label{fig8}
\end{figure}

Turning next to the system of a chain with multiple branches attached at one site, the
most important parameters are the number of branches $M$ and their lengths $l_m$.
The variation of $T(E)$ with the number of branches is illustrated in Figure \ref{fig8}, for
branches of length 2. 
For only a single branch (Figure \ref{fig8}(a)), $T(E)$ shows a single
resonance peak at $E=0$, with a pair of smaller peaks to the sides, separated by a pair
of anti-resonances, at $E=-0.4$ and $E=0.65$. Adding a second branch 
produces a similar $T(E)$ curve (Figure \ref{fig8}(b)), in that the resonance and the
two anti-resonances persist at their previous energies. However, there is a general dampening
of the transmission, resulting in lower $T(E)$ at all energies, except the resonance. This effect
is enhanced with the addition of a third branch (Figure \ref{fig8}(c)), where transmission
is further suppressed, but with the resonance and both anti-resonances preserved. In the more
general situation, where atoms on different branches have different site energies ($\alpha$'s), more
anti-resonances (and, hence, more peaks) appear, but the addition of more branches again has
the overall effect of lowering transmission. Thus, we make the general observation that more
branches is inhibitive to transmission. 

\begin{figure}[h]
\includegraphics[width=12cm]{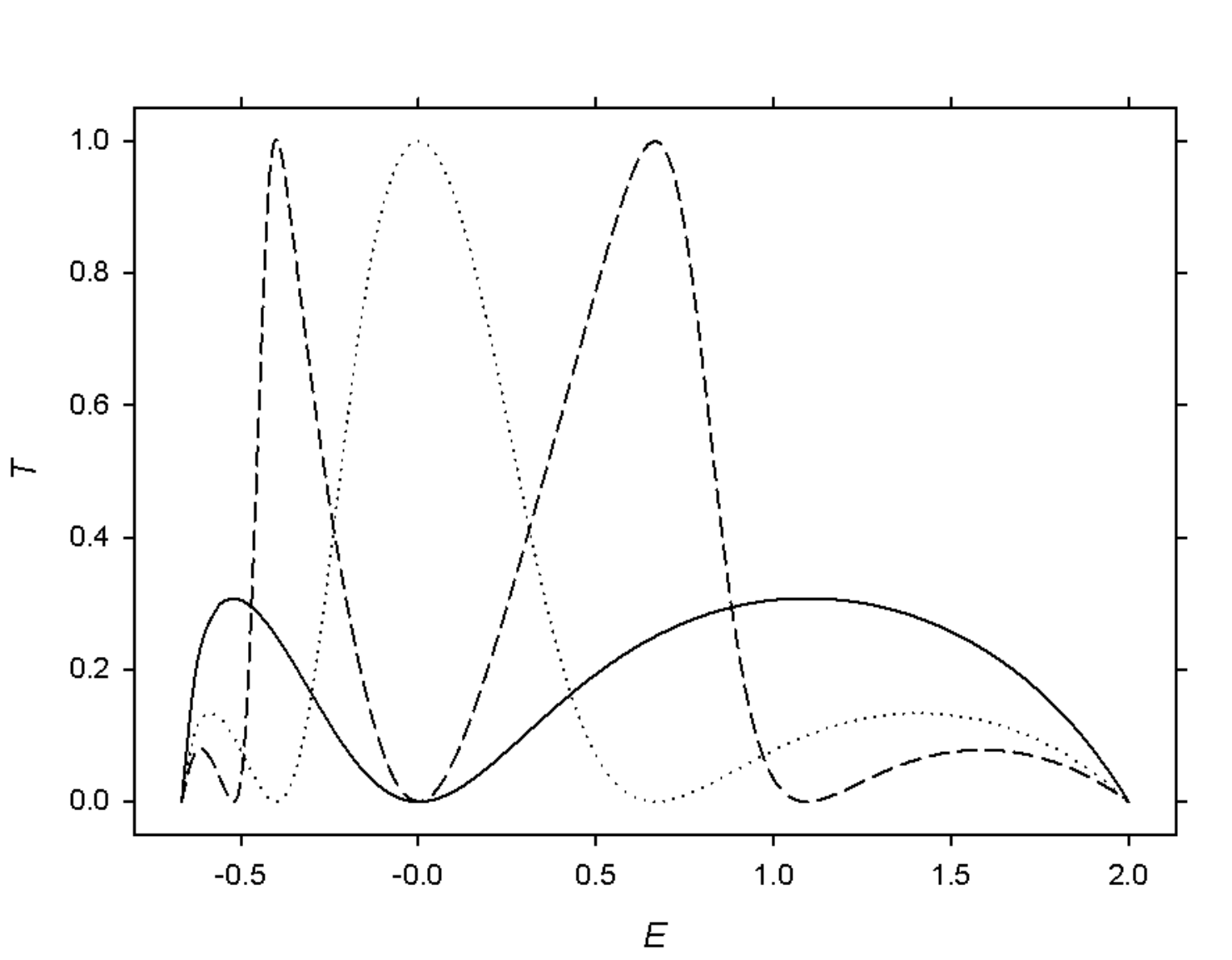}
\caption{$T(E)$ for chain with $M=3$ branches, all of length $l_m =$ (a) 1 (solid curve), (b) 2 (dotted), (c) 3 (dashed).}
\label{fig9}
\end{figure}

The dependence of $T(E)$ on the length of branches is shown in Figure \ref{fig9}, where the
number of branches is taken to be 3. 
For branches of length 1 (i.e., they are individual atoms), 
the $T(E)$ curve (Figure \ref{fig9}(a)) shows relatively low transmission at all energies, with a
single anti-resonance at $E=0$ separating two small peaks. Increasing the lengths of the branches
to 2 produces a markedly different structure (Figure \ref{fig9}(b)), in which a central peak, with
a resonance at $E=0$, appears, with 2 much smaller peaks to the sides, and a pair of anti-resonances
separating them from the central peak. A further increase of the branch lengths to 3 (Figure \ref{fig9}(c))
produces a third anti-resonance, with the 2 central peaks now becoming resonances, and the 2 outside
peaks showing even lower transmission. This trend continues as the chains are further lengthened, with
additional anti-resonances appearing and all the resulting peaks, except the 2 outer ones, showing resonances.
As an individual branch is lengthened (graphs not shown), the number of anti-resonances increases in a pattern 
similar to that noted earlier, in the discussion of Figure \ref{fig6}.

Much of the basic structure of this system can be illuminated by, as before, analyzing
the number of anti-resonances, as illustrated in Figures \ref{fig8} and \ref{fig9}.
As with the first system, anti-resonances are calculated as singularities of 
$\bar{\alpha}_a - \alpha$, now using (\ref{eq16}). The main result from earlier is
still valid, namely, that the maximum number of anti-resonances arising from a branch equals the length $l_m$
of the branch. Thus, for a system with $M$ branches, the maximum number of anti-resonances in
the system equals $l_1 + \cdots + l_M$. However, for the parameter values used here, where each
branch is assumed to have identical atoms, so that corresponding parameters ($\alpha_{am}$, etc,) 
are equal, the number of anti-resonances may turn out to be less than the maximum. In particular,
if the branches all have equal length, as is the case in Figures \ref{fig8} and \ref{fig9}, then the 
anti-resonances arising from one branch coincide with those from the other branches, so that the
number of anti-resonances equals the common branch length $l_m$. But if the branch lengths are
unequal, then typically the anti-resonances from different branches do not coincide, and there is
no reduction in total number from the maximum.

\begin{figure}[h]
\includegraphics[width=12cm]{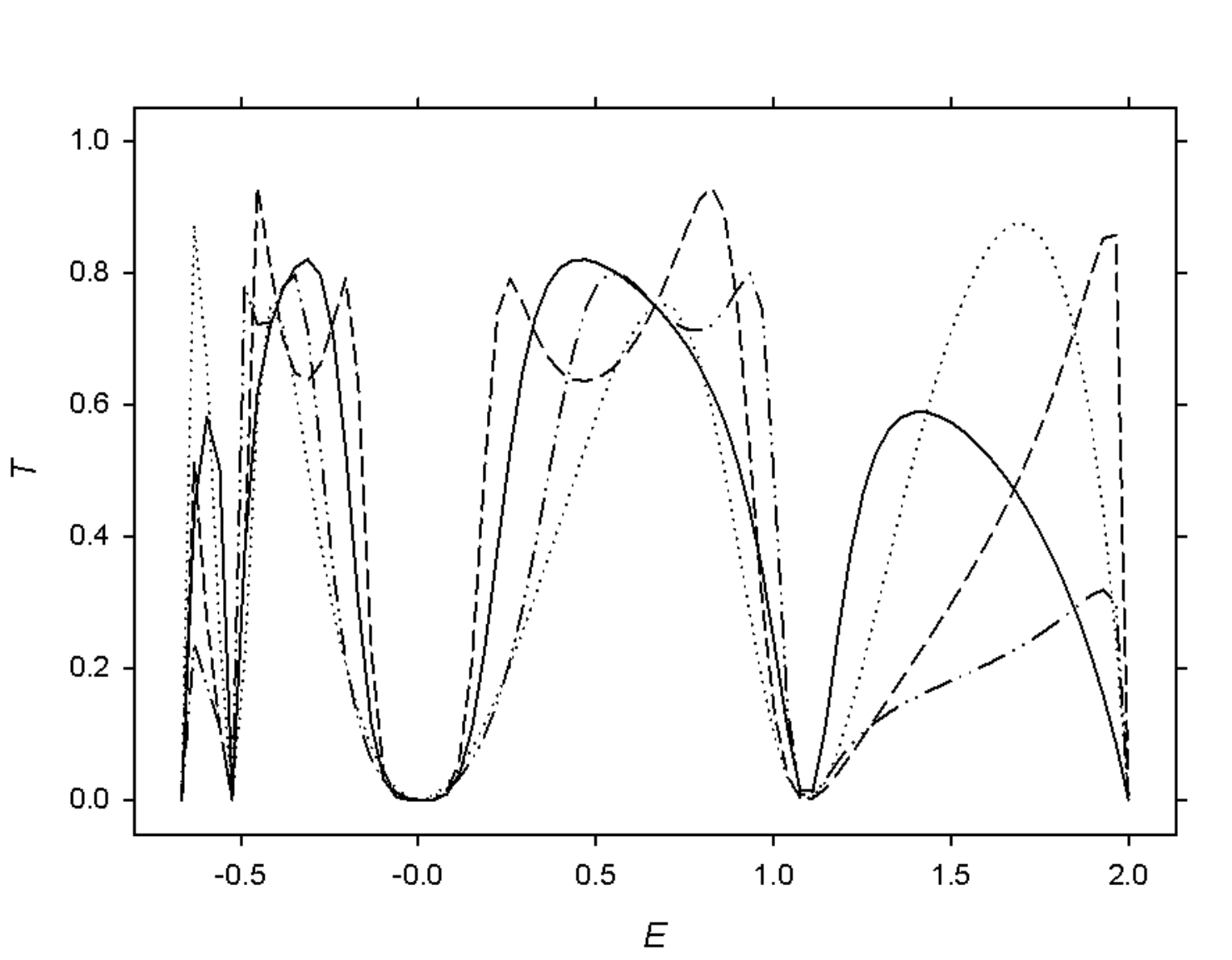}
\caption{$T(E)$ for chain with 2 branches, of lengths $l_a=1$ and $l_b=3$, and at separation $d=$ (a) 1 (solid curve), 
(b) 2 (dotted), (c) 3 (dashed), (d) 4 (dash-dotted).}
\label{fig10}
\end{figure}

Lastly, turning to the system of a chain with a pair of branches, at separate sites, the significant parameters
are the separation $d$ between the two branches, and their lengths $l_a$ and $l_b$. The variation of $T(E)$
with the separation between branches is shown in Figure \ref{fig10}, for branches of fixed lengths 1 and 3. 
(It
should be noted that $T(E)$ is invariant to interchange of the lengths $l_a$ and $l_b$.) When the branches are 
attached to adjacent sites in the chain, corresponding to $d=1$ (Figure \ref{fig10}(a)), the $T(E)$ curve shows
a structure which is quasi-symmetric about $E=0$. (In the case where there is no overlap, the curve
is precisely symmetric.) There are 3 anti-resonances, including one at $E=0$, which divide the band region
into 4 zones of transmission, albeit with no resonances. As the separation increases  (Figure \ref{fig9}(b-d)),
the number and positions of the anti-resonances do not change, leading to a very similar graph for $d=2$, but
to noticeably different ones for $d=3$ and 4, where the inner pair of peaks exhibit some rearrangement
to produce double-peaked structures, while the outer pair are greatly diminished, at least for $d=4$. For further
increases in separation (graphs not shown), the three anti-resonances remain pinned at their energy values,
while the transmittivity continues to be dominated by the inner pair of sub-bands, which gradually add extra
peaks within each zone.

\begin{figure}[h]
\includegraphics[width=12cm]{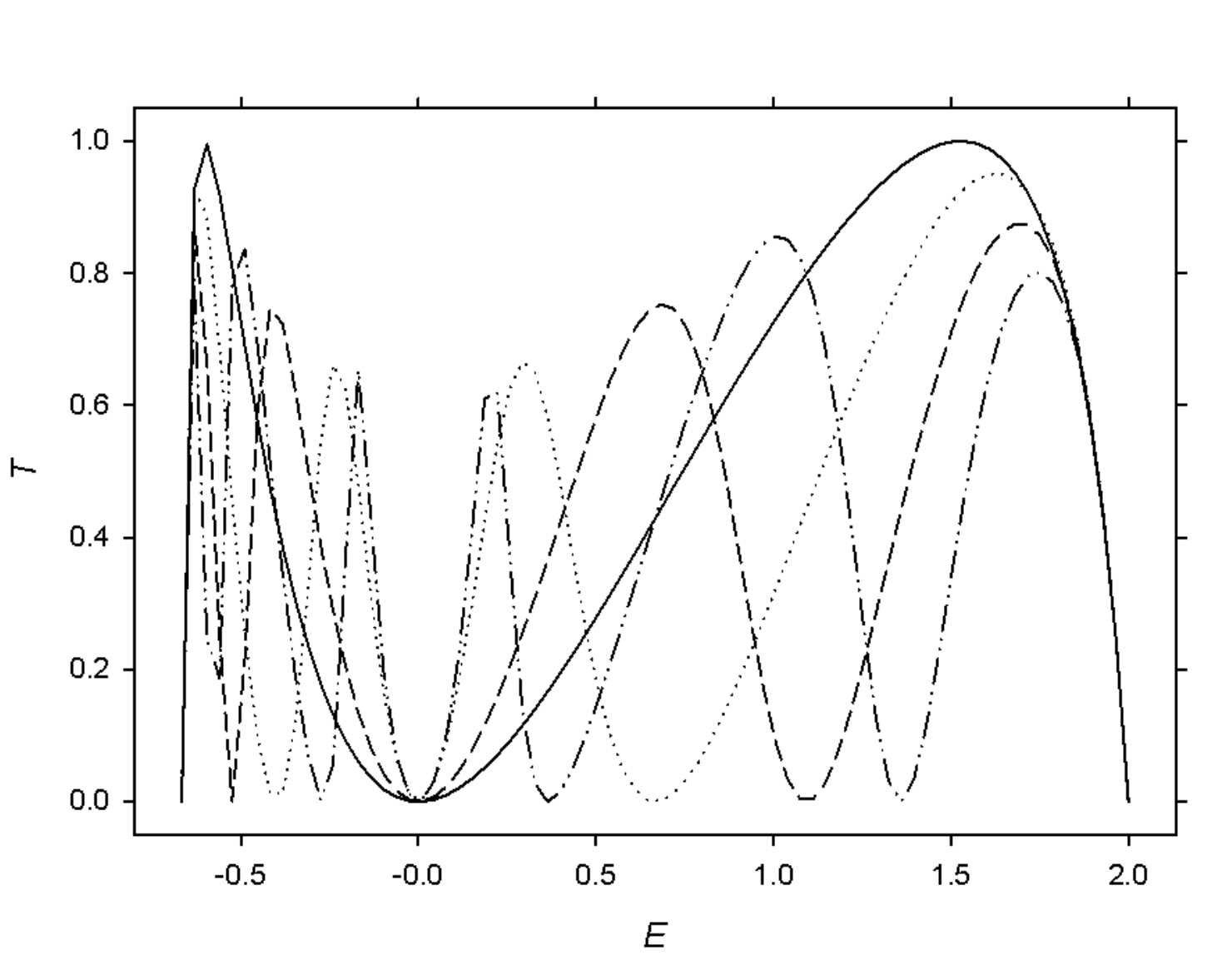}
\caption{$T(E)$ for chain with 2 branches, at separation $d=2$, and lengths $l_a=1$ and $l_b=$ (a) 1 (solid curve), 
(b) 2 (dotted), (c) 3 (dashed), (d) 4 (dash-dotted).}
\label{fig11}
\end{figure}

The variation of $T(E)$ with branch length is exemplified by Figure \ref{fig11}, where the length $l_b$ of 
one branch is varied, while the other $l_a=1$ is held constant, with fixed separation $d=2$. 
For the shortest
branch $l_b=1$ (Figure \ref{fig11}(a)), there is a single anti-resonance at $E=0$ separating two bands
of transmission, with resonances at about $E=-0.5$ and $E=1.5$. The asymmetry, produced by the presence
of overlap, distorts the $T(E)$ curve, so that the part of the band most amenable to transmission is at the
higher energies. As the branch is lengthened to $l_b=2$ (Figure \ref{fig11}(b)), each region of transmission 
is split into two by the creation of a pair on anti-resonances, while the heights of all 4 peaks are lowered, so as
to remove both resonances. The result, then, is an overall weakening of the transmittivity of the wire, except
at certain energies. Lengthening the branch again to $l_b=3$ (Figure \ref{fig11}(c)) produces a graph very
similar to the previous one, with the same number of anti-resonances and peaks, but with some shifting of
their energies (except, noticeably, the anti-resonance at $E=0$). However, a further lengthening of the 
branch to $l_b=4$ creates a splitting of the two inner peaks (Figure \ref{fig11}(d)), for a total of 6, separated 
by 5 anti-resonances (including the fixed one at $E=0$). The net effect is an overall lowering of the 
transmittivity, except at and near the peaks. The structure in the graph for $l_b=4$ persists at
$l_b=5$ (not shown), but when $l_b=6$, there is a further splitting of the innermost pair of peaks, 
accompanied by the appearance of an extra pair of anti-resonances. Thus, we see a pattern that
is reminiscent of that observed with the lengthening of branches in the two earlier systems. 

An analysis of the anti-resonances follows similar lines to that presented for the other two systems, but
now utilizing (\ref{eq17}). First, we note that the number and positions of anti-resonances
are independent of the separation $d$ between the two branches, as can be seen in Figure \ref{fig10}.
In general, there are anti-resonances associated with each branch, the number being equal to
the length of the branch, so that the maximum number of anti-resonances is $l_a + l_b$. However,
as in the earlier systems, taking corresponding parameters on the two branches to be equal, typically
reduces the actual number of anti-resonances. Specifically, this reduction occurs when $l_a$ and $l_b$
are both odd, which results in $E= \alpha_a = \alpha_b$ being an anti-resonance energy for both 
branches, thus reducing the total number, by  1, to $l_a + l_b-1$. For certain other combinations
of lengths, the total number of anti-resonances can also be reduced. 
A special situation occurs when $l_a=l_b$, with corresponding 
parameters equal, so that the two branches are identical. In this case, the number of 
anti-resonances equals the common length $l_a$. These considerations explain the pattern
of anti-resonances seen in Figure \ref{fig11}.

\section{Conclusion}

Starting from the bare atomic wire, more complicated electronic systems can be built, by
adding one or more side branches, at one or more sites, along  the wire. By examining
these different structures, we can gain insight into those variations which produce the most
useful properties. The present work utilizes the renormalization-decimation method, whereby
all the side branches, attached to a particular site, were reduced to an impurity replacing the
original atom at that site. Subsequently, the transmission probability $T(E)$ through the wire was
calculated by means of the established Lippmann-Schwinger technique.

In all of the three systems examined, the key to understanding the $T(E)$ curve lies in
 the number and positions of anti-resonances and resonances (or smaller maxima). These
in turn can be varied and controlled by means of the system's main parameters, such as the length
of a side branch, thus indicating that the wire can be constructed to have a fairly specific $T(E)$
profile, through a judicious choice of the attached structures.

It is interesting to note that the results of the present study indicate that the renormalization-decimation
technique may be a useful tool in the area of chemisorption. For example, it could be used to
investigate the chemisorption properties of a molecule adsorbed onto an atomic wire, by treating
the admolecule similarly to the side-structures of this paper. In such a case, the admolecule can
be renormalized down to a single impurity in the wire, thus simplifying the chemisorption 
calculation.

\section{Keywords}

Branched atomic wires, molecular electronics, renormalization-decimation technique, semi-empirical calculations

\end{document}